\documentclass[prl,showpacs,twocolumn,amsmath]{revtex4}
\usepackage{graphicx}

\providecommand{\MSbar }{\ensuremath{ \overline{\rm MS} }}

\begin{document}
%%%%%%%%%%%%%%%%%%%%%%%%%%%%%%%%%%%%%%%%%%%%%%%%%%%%%%%%%

\title{Universality of soft and collinear factors in hard-scattering 
factorization} 

\author{John C. Collins}
\affiliation{Penn State University, 
104 Davey Lab, University Park PA 16802, U.S.A.
}
\author{Andreas Metz}
\affiliation{Institut f{\"u}r Theoretische Physik II, 
Ruhr-Universit{\"a}t Bochum, 44780 Bochum, Germany
}

\date{December 10, 2004}

%%%%%%%%%%%%%%%%%%%%%%%%%%%%%%%%%%%%%%%%%%%%%%%%%%%%%%%%%%%%%%%%%%%%%%%%%%%
\begin{abstract}
  Universality in QCD factorization of parton densities, fragmentation
  functions, and soft factors is endangered by the process dependence
  of the directions of Wilson lines in their definitions.  We find a
  choice of directions that is consistent with factorization and that
  gives universality between $e^+e^-$-annihilation, semi-inclusive
  deep-inelastic scattering, and the Drell-Yan process.  Universality is only
  modified by a time-reversal transformation of the soft function and
  parton densities between Drell-Yan and the other processes, whose
  only effect is the known reversal of sign for $T$-odd parton
  densities like the Sivers function.  The modifications of the
  definitions needed to remove rapidity divergences with light-like
  Wilson lines do not affect the results.
\end{abstract}


\maketitle

%%%%%%%%%%%%%%%%%%%%%%%%%%%%%%%%%%%%%%%%%%%%%%%%%%%%%%%%%%%%%%%%%%%%%%%%%%%
%\section{Introduction}
\paragraph{Introduction}

Much of the predictive power of QCD is provided by universality of the
non-perturbative functions in factorization theorems for hard
processes.  These functions are parton densities, fragmentation
functions, etc.  Whereas the perturbative parts of factorization
formulae can be usefully estimated from first principles by weak
coupling methods, the non-perturbative functions cannot.  Universality
is the process independence of these functions.  It allows them to be
measured from a limited set of reactions, and then used to predict
other reactions, with the aid of factorization and perturbative
calculations.

However, recent developments 
\cite{collins_02,BHS,BHS.DY,boer_03,bomhof_04} show
that universality is endangered.  For example, the Sivers function
\cite{sivers_90}
--- the transverse single-spin asymmetry of a parton density --- 
changes sign \cite{collins_02,BHS.DY} between the Drell-Yan process and
deep-inelastic scattering.  This is because the directions of the
Wilson lines necessary for a gauge-invariant definition of a parton
density depend on whether collinear and soft interactions are before
or after the hard scattering.  Important current experimental work 
\cite{expt} addresses the associated physics issues.

Although for parton densities time reversal relates the two
definitions \cite{collins_02}, the situation is not so clear in
general.  For example, a fragmentation function involves a
semi-inclusive sum over out-states:
\begin{equation}
  \sum_X |A,X, \text{out}\rangle ~ \langle A,X, \text{out}| .
\end{equation}
Time-reversal converts these to in-states, and therefore does not
prove universality with the obvious, process-dependent directions for
the Wilson lines \cite{boer_03} --- although the non-universality did
not occur in a one-loop model calculation \cite{metz_02}.  Moreover,
for hadron production in hadron-hadron collisions, Bomhof, Mulders,
and Pijlman \cite{bomhof_04} found a jungle of Wilson lines whose
universality properties are far from clear;
see also the comments of Brodsky, Hwang, and Schmidt \cite{BHS.DY}.

Therefore in this paper we carefully re-examine the arguments about
Wilson lines to discover the true limits of universality, if any.  
The issues particularly concern processes that need 
transverse-momentum-dependent (TMD)
partonic functions,
and we consider three such processes: 
(a) $e^+e^-$ annihilation with detected almost
back-to-back hadrons \cite{collins_81}, (b) semi-inclusive
deep-inelastic scattering (SIDIS) with measured transverse momentum
for an outgoing hadron \cite{meng,ji_04}, and (c) the Drell-Yan process with
measured transverse momentum \cite{collins_85a}.  For the first
process, Collins and Soper showed a factorization theorem more than
two decades ago \cite{collins_81}.  But the paper \cite{collins_85a}
extending the \emph{statement} of factorization to the Drell-Yan
process claimed no proof.

Our new methods show, in addition to the time-reversal-modified
universality \cite{collins_02} of parton densities, that:
(1) TMD fragmentation functions are universal between $e^+e^-$
  annihilation and SIDIS.
(2) The soft factor is universal between all three processes.
(3) Universality arguments hold even with a redefinition of the
  non-perturbative functions needed to remove the divergences due to
  light-like Wilson lines.
Our arguments delimit the process-dependent choices of direction that
are compatible with factorization, and for individual processes the
choice is wider than previously used \cite{boer_03,ji_04,collins_00}.

The central technical issue is that a proof of factorization requires
an appropriate deformation \cite{collins_81a} of momentum contours out
of the ``Glauber region''.  Allowed directions of the Wilson lines are
those compatible with the contour deformation.  The possible
directions of contour deformation are determined by the space-time
location of soft and collinear interactions relative to the hard
scattering.  Hence a careful analysis of one-gluon corrections, which
forms the bulk of our work in this letter, should be sufficient to determine
the directions.

%%%%%%%%%%%%%%%%%%%%%%%%%%%%%%%%%%%%%%%%%%%%%%%%%%%%%%%%%%%%%%%%%%%%%%%%%%%
%\section{Factorization and Wilson lines}
\paragraph{Factorization and Wilson lines}

Factorization into hard, collinear and soft factors results
essentially from a corresponding set of momentum regions.  Individual
graphs for a cross section have a range of possible leading regions of
loop-momentum space, and we will follow the approach of Collins and
Hautmann \cite{collins_00} by using a subtractive method that
implements a decomposition of graphs by possible regions.  Full
details of the subtractive method to all orders have not been worked
out explicitly, but it should give factorization after a sum over
graphs and regions.  A Ward identity argument is needed to disentangle
otherwise coupled factors, and it results in simple Wilson lines in
the gauge-invariant operator definitions of the factors.

%%%%%%%%%%%%%%%%%%%%%%%%%%%%%%%%%%%%%%%%%%
%\subsection{Electron-positron annihilation}
\paragraph{Electron-positron annihilation}

%%%%%%%%%%%%%%%%
\begin{figure}
\centering
\includegraphics[scale=0.65]{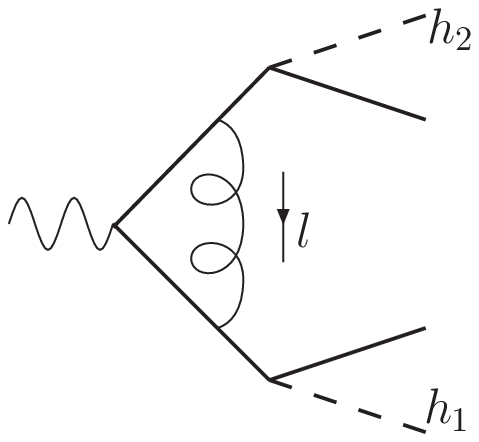}~~
\includegraphics[scale=0.65]{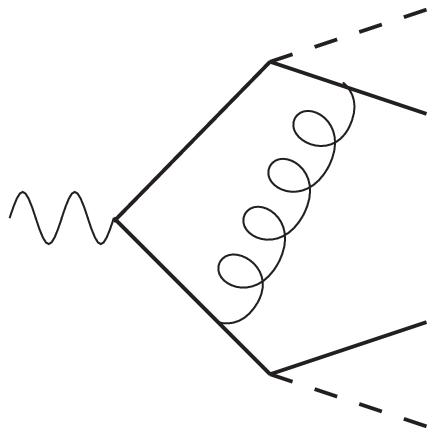}
\caption{Sample virtual one-loop diagrams for $e^+ e^-$
  annihilation, with fragmentation considered in a spectator
  model. 
\label{f:epem}}
\end{figure}
%%%%%%%%%%%%%%%%

We first consider $e^+e^-$ annihilation at large $Q$ with two detected
hadrons $h_1$ and $h_2$ in almost back-to-back directions, for which some
simple graphs with a virtual gluon are shown in Fig.~\ref{f:epem}.
(The contour deformation issues arise only for virtual gluons.)

\begin{widetext}
For the first graph, we make, as in \cite{collins_00}, a
decomposition into four terms: hard, collinear to $h_1$, collinear
to $h_2$, and soft.  Critical for determining the directions
of the Wilson lines is the soft term:
\begin{equation}
\label{eq:epem.soft}
\begin{split}
  S_{\rm epem} 
 = {}&
  \int \frac{d^nl}{(2\pi)^n} \,
  F_1^+(P_1,-(0,l^-,0_T)) \,
  F_2^-(P_2,(l^+,0,0_T)) \,
  \frac{ i l^-l^+ }{ l^2 + i\epsilon }
\\
  &\quad \left[
     \frac{1}{ (-l^- +i\epsilon) \, (l^+ +i\epsilon) }
     - \frac{1}{ (-l^- +i\epsilon) \, (l^+ - C_Al^- +i\epsilon) }
     - \frac{1}{ (-l^- + C_Bl^+ +i\epsilon) \, (l^+ +i\epsilon) }
  \right]
 + \mbox{\MSbar{} counterterm}.
\end{split}
\end{equation}
Here we use light-front coordinates in the center-of-mass with
the dominant components of the momenta of $h_1$ and $h_2$ being
$P_1^+$ and $P_2^-$.
\end{widetext}

In the $h_1$ and $h_2$ parts of the graph we have retained only the
$-$ and $+$ components of the gluon momentum $l$, and we have picked
out the dominant components of the current to which the gluon couples.
We have inserted factors of $l^-$ and $l^+$ to allow Ward identities
to be used and then compensated this in the first term in brackets by
dividing by $l^-$ and $l^+$.  With this first term we obtain a good
approximation to the original graph in the soft region, provided that
we make a suitable deformation out of the Glauber subregion of the
soft region.  The Glauber region is where $|l^+l^-| \ll l_T^2$.  The
second and third terms are counterterms, to cancel the
divergences at large positive and negative gluon rapidity that would
arise if only the first term were used.  

The counterterms have arbitrary positive parameters $C_A$ and $C_B$ of
order $Q^2/m^2$ (with $m$ being a soft scale),
they are power suppressed in the soft region, and they
allow a Wilson line definition of the soft factor.  The arbitrariness
of $C_A$ and $C_B$ is exploited by the use of the Collins-Soper
\cite{collins_81} equation, which controls the $C_A$ and $C_B$
dependence of the soft and collinear factors, and so enables
predictions to be made.

The contour deformation must not cross the final-state quark
poles in the original graph.  We choose to deform symmetrically out of
the Glauber region: $\Delta l^+ = iw$, $\Delta l^- = - iw$, where $w$ is a
suitable positive function of the real parts of $l^\mu$.  Compatibility
with this deformation determines uniquely that the light-like
Wilson-line denominators are $-l^-+i\epsilon$ and $l^++i\epsilon$, and that the
signs of the $C_A$ and $C_B$ terms relative to their $i\epsilon$s are as
shown.  But it does not determine whether the counterterm lines are
space-like or time-like.  A legitimate possibility not noticed in
\cite{collins_00} is that one or both could be
time-like: $l^+ +C_Al^- -i\epsilon$, $-l^- -C_Bl^+ -i\epsilon$.

There is also an ultra-violet divergence at large $l_T$ which we
remove by ordinary renormalization.  Neither the rapidity divergences
nor the UV divergence affect the validity of the soft approximation
in the soft region.

In coordinate space, both light-like Wilson lines are future pointing,
as is intuitively natural.  They approximate a fast-moving quark and a
fast-moving antiquark as seen by a slow gluon.  (However, the
non-light-like lines in the counterterms are past-pointing, not so
intuitively.)

We choose space-like lines in the counterterms because they are
compatible with one of the possibilities for the counterterms in
SIDIS, and so they allow a proof of universality.
Furthermore, exact properties of matrix elements of Wilson lines are
simpler when the gluon fields are at space-like separation and
hence all commute.

Another possible change is to use an asymmetric contour deformation,
such as we choose in SIDIS, i.e., primarily on $l^+$ only or on $l^-$.
But it can be shown that the only extra resulting cases for the
$i\epsilon$ prescriptions violate the charge-conjugation relationships
between fragmentation for quarks and antiquarks.  They would therefore
remove predictive power from factorization.

In the method of Ji, Ma, and Yuan \cite{ji_04} there are no
counterterms; instead they use slightly non-light-like lines to cutoff
the rapidity divergences.  Their square-bracket factor would be
\begin{equation}
\label{eq:epem.nll}
   \frac{1} 
   { (l^+/C_A - l^- +i\epsilon) \, (-l^-/C_B + l^+ +i\epsilon) }.
\end{equation}
The Wilson lines are actually future-pointing, since the denominators
are on opposite sides of the graph compared with the corresponding
past-pointing space-like lines in Eq.\ (\ref{eq:epem.soft}).  Some
differences with our formulation are inessential power-law
corrections.  But there are also different leading-power
contributions to the soft and collinear factors.  We remark without
proof that these only occur at large transverse momentum for the gluon
and therefore amount to a legitimate scheme change.  The subtractive
method provides simpler calculations and a cleaner proof of
universality.

Once the soft term is fixed, definite prescriptions for the collinear
and hard terms follow, just as in \cite{collins_00}, so we will not
present them here.

There are many other graphs for the process, both with different
connections of the gluon, as in Fig.\ \ref{f:epem}, and with
arbitrarily many other lines.  In all leading regions, the collinear
parts are in the final state, so we can continue applying the same
prescription for the contour deformations and for the counterterms.
Of course, to use Ward identities we must use the same
prescription everywhere.  So we have determined all the Wilson lines.

%%%%%%%%%%%%%%%%%%%%%%%%%%%%%%%%%%%%%%%%%%
%\subsection{Semi-inclusive DIS}
\paragraph{Semi-inclusive DIS}

\begin{figure}
\centering
\includegraphics[scale=0.65]{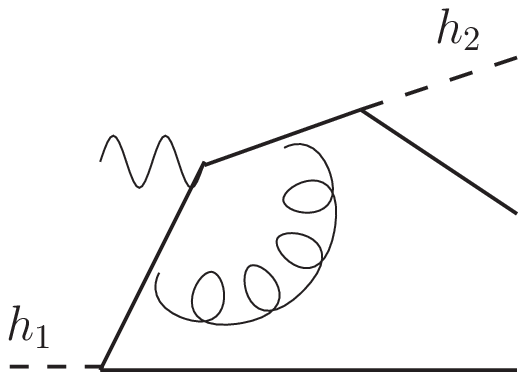}~~
\includegraphics[scale=0.65]{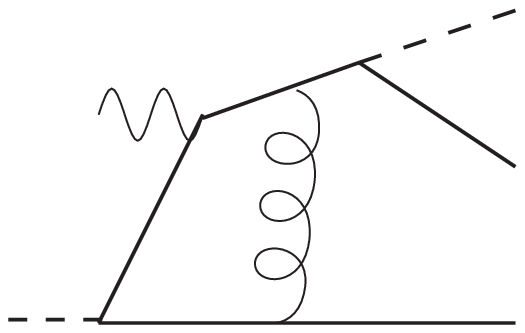}
\caption{Sample one-loop diagrams for SIDIS.
\label{f:dis}}
\end{figure}

We now consider similar graphs for SIDIS, as in Fig.\ \ref{f:dis}.  As
concerns the contour deformation from the Glauber region, the primary
difference is that in the $h_1$ part of the graph the flow of $l$
relative to collinear momenta is reversed.  So we have denominators
like $(k_1+l)^2-m^2+i\epsilon$ instead of $(k_1-l)^2-m^2+i\epsilon$.
This suggests reversing the contour deformation on $l^-$, to give
$\Delta l^+ = iw$, $\Delta l^- = iw$.  Then, following
\cite{ji_04}, we might reverse the relative signs of $l^-$ and
$i\epsilon$ compared with Eq.\ (\ref{eq:epem.soft}), to get a
square-bracket factor
\begin{equation}
\label{eq:dis.soft.T}
\begin{split}
     \frac{1}{ (-l^- -i\epsilon) \, (l^+ +i\epsilon) }
     - \frac{1}{ (-l^- -i\epsilon) \, (l^+ + C_Al^- +i\epsilon) }
\\
     - \frac{1}{ (-l^- - C_Bl^+ -i\epsilon) \, (l^+ +i\epsilon) }.
\end{split}
\end{equation}
The counterterm lines are now time-like.

%%%%%%%%%%%%%%%%
%%%%%%%%%%%%%%%%

Corresponding time-like lines must also appear in the fragmentation
function, so that compared with $e^+e^-$-annihilation, the definitions
of both the fragmentation function and the soft factor are different,
and we lose manifest universality.  Now both of $C_A$ and $C_B$ are
large, so that we can in fact use space-like lines with our symmetric
contour deformation (and similarly in the Ji, Ma and Yuan version):
\begin{equation}
\label{eq:dis.soft}
\begin{split}
     \frac{1}{ (-l^- -i\epsilon) \, (l^+ +i\epsilon) }
     - \frac{1}{ (-l^- -i\epsilon) \, (l^+ - C_Al^- -i\epsilon) }
\\
     - \frac{1}{ (-l^- + C_Bl^+ +i\epsilon) \, (l^+ +i\epsilon) }.
\end{split}
\end{equation}
But even this still differs from the version for
$e^+e^-$-annihilation, so we cannot directly deduce universality.

Moreover, in the second graph of Fig.\ \ref{f:dis} the symmetric
contour deformation is blocked by a trap between initial- and
final-state poles in the target part of the graph.  When transverse
momenta are of order $m$, the deformation on $l^-$ is limited to
$m^2/Q$, not enough to get out of the Glauber region.  So, as in the
proof of factorization for diffractive DIS \cite{collins_97}, we
should deform primarily on $l^+$, away from the pole(s) in
the outgoing struck quark and its jet; this typically makes $l$ collinear to
the target, rather than merely soft.  Only small deformations on $l^-$
are necessary to avoid the target-side poles both in the original
graph and in the Wilson-line approximations.

We can now use exactly the same square-bracket factor as in
$e^+e^-$-annihilation.  Since the same direction of contour
deformation can be applied generally, for the Glauber regions for all
graphs, the Wilson lines in the soft and fragmentation factors are the
same as in $e^+e^-$-annihilation.  Thus the soft and fragmentation
factors are universal between SIDIS and $e^+e^-$-annihilation.
Space-like counterterm denominators, like $l^+ - C_Al^- +i\epsilon$
are preferred here, since the large positive imaginary part of $l^+$
assists rather than hinders the deformation of $l^-$ from an
unphysical pole.

%%%%%%%%%%%%%%%%%%%%%%%%%%%%%%%%%%%%%%%%%%
%\subsection{Drell-Yan process}
\paragraph{Drell-Yan process}

%%%%%%%%%%%%%%%%
\begin{figure}
\centering
\includegraphics[scale=0.6]{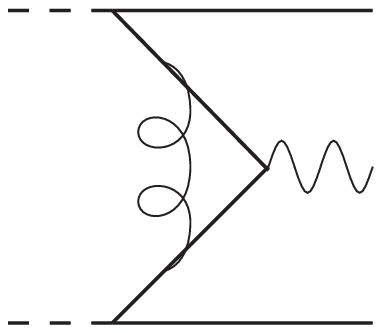}~~
\includegraphics[scale=0.6]{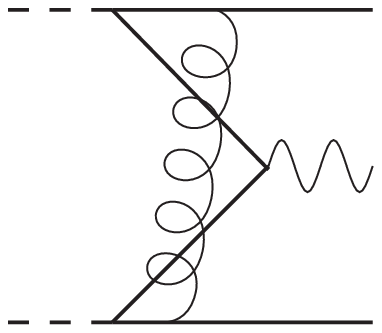}
\caption{Sample one-loop diagrams for the Drell-Yan process.
\label{f:dy}}
\end{figure}
%%%%%%%%%%%%%%%%

In Fig.\ \ref{f:dy}, we show some graphs for the Drell-Yan process.
In the first graph, the gluon attaches to two initial-state
lines, so we use a contour deformation opposite to that in
$e^+e^-$ annihilation.

But other graphs trap the contour against final-state poles in the
target parts of the graphs.  Now to prove factorization
\cite{collins_85_88} one can deform away from initial-state poles.
Crossing target-related final-state poles produces extra
non-factorizing terms, but these cancel by unitarity, after a sum over
all hadronic states in the inclusive cross section.  The argument does
not depend on the transverse momentum of the lepton pair.

Therefore we find that we can use initial-state Wilson lines for the
parton densities and for the soft factor.  The time reversal argument
of \cite{collins_02} applies to all these objects, since they all
involve only matrix elements of the form $\langle \psi |A_1 A_2
|\psi\rangle$, where $A_1$ and $A_2$ are operators and $\psi$ labels a
vacuum or a one-particle state.  Such states are the same no matter
whether they are in- or out-states, so the transformation by
time-reversal leaves them unaffected.

This extends the exact universality of parton densities and soft
factors to the Drell-Yan process, with the exception of ``$T$-odd''
parton densities, which reverse sign \cite{collins_02}, as is already
known.  Our proof now includes Wilson-line factors that implement
\cite{collins_03} the cancellation of rapidity divergences.

%%%%%%%%%%%%%%%%%%%%%%%%%%%%%%%%%%%%%%%%%%%%%%%%%%%%%%%%%%%%%%%%%%%%%%%%%%%
%\section{Conclusions and discussion}
\paragraph{Conclusions and discussion}

We have shown universality of fragmentation functions, soft factors,
and parton densities between $e^+e^-$-annihilation, semi-inclusive
deep-inelastic scattering, and the Drell-Yan process.  This applies both to the
basic definitions with light-like Wilson lines and to the correct
definitions with removal of rapidity divergences.  Regulator lines
should be space-like.  In the Drell-Yan process the lines are reversed compared
with the other processes we considered, but time-reversal relates them
to the functions for other processes, with the usual reversal of sign
for the Sivers function and other ``$T$-odd'' parton densities.

This eliminates missing elements in the Collins-Soper-Sterman
formalism \cite{collins_85a} for the Drell-Yan process.

The method of Ji, Ma, and Yuan \cite{ji_04} uses non-light-like Wilson
lines instead of counterterms for removing rapidity divergences.
Within this method we find universality if the fragmentation function
and the soft factor in SIDIS have future-pointing space-like Wilson
lines, contrary to the choice made by these authors.  

Since we need definitions of TMD parton densities that differ from the
most obvious ones that use light-front quantization in light-front
gauge, we agree with the conclusion of Brodsky et
al.\ \cite{SF.not.prob} that parton densities are not literally
probability densities.  However, our reasoning is different, and
builds on the much earlier work of Collins and Soper
\cite{collins_81,collins_82}.

Our work needs extension to hard hadron-production processes
in hadron-hadron collisions.  It is not obvious that the extension
will succeed.  Factorization for these processes is at present an
unproved conjecture, at least for cases where
transverse-momentum-dependent parton densities and fragmentation
functions are needed.

Even for conventional hadron-hadron-to-hadron factorization with
ordinary, integrated parton densities, there is no proof in the
literature which correctly treats the Glauber region, as far as we
know. The factorization proofs of Collins, Soper and Sterman
\cite{collins_85_88} and Bodwin \cite{bodwin_85} are only for the
Drell-Yan process.  A critical re-examination is needed here.

%%%%%%%%%%%%%%%%%%%%%%%%%%%%%%%%%%%%%%%%%%%%%%%%%%%%%%%%%%%%%%%%%%%%%%%%%%%%%%%%
\begin{acknowledgments}
This work has been supported in part by the U.S. Department of Energy under
grant number DE-FG02-90ER-40577, and by the Sofia Kovalewskaya Programme of
the Alexander von Humboldt Foundation.
\end{acknowledgments}

%%%%%%%%%%%%%%%%%%%%%%%%%%%%%%%%%%%%%%%%%%%%%%%%%%%%%

%%%%%%%%%%%%%%%%%%%%%%%%%%%%%%%%%%%%%%%%%%%%%%%%%%%%%%%%%%%%%%%%%%%%%%%%%%%%%
\end{document}